\documentclass[aps,prbbib,twocolumn,epsf]{revtex4}
\usepackage{graphicx}
\usepackage{bm}
\usepackage{setspace}
\begin{document}
\title{Current-induced torques in the presence of spin-orbit coupling}
\author{Paul M. Haney and  M. D. Stiles}
\affiliation{Center for Nanoscale Science and Technology, National
Institute of Standards and Technology, Gaithersburg, Maryland
20899-6202, USA }

\begin{abstract}
In systems with strong spin-orbit coupling, the relationship between
spin-transfer torque and the divergence of the spin current is
generalized to a relation between spin transfer torques, total
angular momentum current, and mechanical torques.  In ferromagnetic
semiconductors, where the spin-orbit coupling is large, these
considerations modify the behavior of the spin transfer torques. One
example is a persistent spin transfer torque in a spin valve: the
spin transfer torque does not decay away from the interface, but
approaches a constant value.  A second example is a mechanical
torque at single ferromagnetic-nonmagnetic interface.
\end{abstract}

 \maketitle

\noindent {\em Introduction}--- Since the prediction
\cite{bergerdw,slonc,berger} of spin transfer torques in
non-collinear ferromagnetic metal circuits, they have been the
subject of extensive research \cite{ralph,miltat}. The possibility
of using spin transfer torque to improve the commercial viability of
magnetic random access memory (MRAM) \cite{katine}, and the rich
non-equilibrium physics involved establish the topic as one of
practical and fundamental interest.  These torques arise from the
exchange interaction between non-equilibrium, current-carrying
electrons and the spin-polarized electrons that make up the
magnetization.  In systems where the spin-orbit coupling is weak,
the torque on the magnetization can be computed from the change in
the spins flowing through the region containing the magnetization.
This relation is a consequence of conservation of total spin.  Here,
we consider systems in which the spin-orbit coupling cannot be
neglected (and hence total spin is no longer conserved).

In systems where spin angular momentum is not conserved, the
relationship between the spin transfer torque and the flow of spins
needs to be generalized. Conservation of {\it total} angular
momentum implies that mechanical torques on the lattice of the
material accompany changes in the magnetization
\cite{richardson,einstein}. This effect has been used for decades to
measure the $g$-factor of metals.  More recent theoretical
\cite{mohanty,kovalev} and experimental \cite{guiti} work considers
the current-induced mechanical torques present at the interface of a
ferromagnet and non-magnet, similar in spirit to the spin transfer
torques on the magnetization present in spin valves.

In this article we develop a theory for current-induced torques
(both spin transfer torques and mechanical torques) in systems with
strong spin-orbit coupling, and apply it to a model of dilute
magnetic semiconductors.  We find that by accounting for the orbital
angular momentum of the electrons, we can relate the change in total
angular momentum flow to spin transfer torques and mechanical
torques.  We study two system geometries where these torques play
important roles.  The first is a spin-valve geometry, which is used
to study the features of spin transfer torques in the presence of
spin-orbit coupling.  The second is a single interface between a
ferromagnet and non-magnet, which elucidates the physics underlying
current-induced mechanical torques.

\noindent {\em Formalism} --- We consider a Hamiltonian consisting
of a spin-independent kinetic and potential energy
$H_0=\frac{-\hbar^2\nabla^2}{2m} + V({\bf r})$, an exchange
splitting $\Delta$, and an atomic-like spin-orbit interaction
parameterized by $\alpha$:
\begin{eqnarray}
H &=& H_0 + \frac{\Delta}{\hbar} \frac{\left({\bf M} \cdot \hat {\bf
s}\right)}{M_s}+ \frac{\alpha}{\hbar^2} \left(\hat{\bf L} \cdot
\hat{\bf s}\right),\label{eq:HKL}
\end{eqnarray}
where $\hat{\bf L}$ and $\hat{\bf s}$ are the electron angular
momentum and spin operators, respectively \cite{footnote1}.  The
exchange splitting arises from a magnetization ${\bf M}$, with
magnitude $M_s$.  We treat the magnetization within mean field
theory.

We consider the torque on the magnetization due to electric current
flow.  The spin transfer torque $\bm\tau_{\rm STT}$ at position
${\bf r}$ from electronic states with spin density ${\bf s}({\bf
r})$ is proportional to the component of spin transverse to the
magnetization \cite{nunez}: $\bm\tau_{\rm STT}({\bf r})=\frac{d {\bf
M}({\bf r})}{dt} = \frac{-\Delta}{\hbar^2} \left({\bf M({\bf r})}
\times {\bf s({\bf r})}\right) $. In the absence of spin-orbit
coupling, this torque can be related to the divergence of a spin
current, which offers conceptual and computational simplicity
\cite{stiles}. In the following we analyze how spin-orbit coupling
changes this simple result.  One consequence is an expression for
the mechanical torque $\bm\tau_{\rm lat}$.

We develop an expression for $\bm\tau_{\rm STT}$ by evaluating the
time-dependence of the electron spin and angular momentum densities.
To do so, we adopt a Heisenberg picture of time evolution, and
evaluate $\frac{d \hat{O}({\bf r})}{dt} = \frac{i}{\hbar}\left[ H,
\hat\psi^\dagger({\bf r}) \hat O \hat\psi({\bf r})  \right] ,$ where
$\hat \psi({\bf r})$ is the position operator, for the operators
$\hat O = {\bf \hat s}$, ${\bf \hat L}$.  This procedure leads to
\cite{ralph}:
\begin{eqnarray}
\frac{d{\bf \hat{s}}}{dt} = \nabla\cdot \hat{\bf Q}_{\bf s}({\bf r})
- \hat{\bm \tau}_{\rm STT} + \frac{\alpha}{\hbar^2} \left(\hat{\bf
L} \times \hat{\bf s}\right)\label{eq:dsdt}
\end{eqnarray}
where $ \hat{\bf Q}_{\bf s}({\bf r}) = \hat\psi^\dagger({\bf r})
{\hat {\bf v}}\otimes {\bf {\hat s}} \hat\psi({\bf r})$, and the
velocity operator is given by ${\bf {\hat v}}=
\frac{i\hbar}{2m}\left( \overleftarrow{\nabla} -
\overrightarrow{\nabla} \right)$; here the arrow superscript
specifies the direction in which the gradient acts.  In addition:
\begin{eqnarray}
\frac{d{\bf \hat{L}}}{dt} = \nabla\cdot \hat{\bf Q}_{\bf L}({\bf r})
- \hat{\bm\tau}_{\rm lat} + \frac{\alpha}{\hbar^2} \left(\hat{\bf s}
\times \hat{\bf L}\right) \label{eq:dldt}
\end{eqnarray}
where $\hat{\bf Q}_{\bf L}({\bf r}) = \hat\psi^\dagger({\bf r})
\frac{1}{2} \left( {\hat {\bf v}}{\bf {\hat L}} + {\hat {\bf L}}{\bf
{\hat v}} \right) \hat\psi({\bf r})$ (the product of non-commuting
operators ${\bf {\hat L}}$ and ${\bf {\hat v}}$ is symmetrized).
We've defined $ \hat{\bm\tau}_{\rm lat}({\bf
r})=\frac{i}{\hbar}\hat\psi^\dagger({\bf r})\left[H_0,{\bf {\hat
L}}\right]\hat\psi({\bf r})$, which is nonzero for a potentials
$V({\bf r})$ which break rotational symmetry \cite{footnote2}.

We define a total angular momentum $\hat{\bf J}=\hat{\bf L}+\hat{\bf
s} $, a total angular momentum current $\hat{\bf Q}_{\bf J}=\hat{\bf
Q}_{\bf L}+\hat{\bf Q}_{\bf s}$, and combine Eqs. (\ref{eq:dsdt})
and (\ref{eq:dldt}) to obtain:
\begin{eqnarray}
\frac{d\hat{\bf J}}{dt}- \bm\nabla \cdot
  \hat{\bf Q}_{\bf J}  = - \hat{\bm\tau}_{\rm STT}
  - \hat{\bm\tau}_{\rm lat}  .\label{eq:stt}
\end{eqnarray}
Finally, we take the expectation value of
Eqs.~(\ref{eq:dsdt}-\ref{eq:stt}), replacing operators by densities.
Eq.~(\ref{eq:stt}) is our main formal result. When spin-orbit
coupling is important, the total angular momentum in the conduction
electrons couples both to the magnetization and the lattice.  The
coupling of electron spin to the lattice requires both spin-orbit
coupling and crystal field potential.  The term $\bm\tau_{\rm lat}$
changes the physical picture of spin transfer torque substantially,
as is illustrated by considering Eq.~(\ref{eq:stt}) for a single
bulk eigenstate: $\frac{d{\bf J}}{dt}$ and $\nabla \cdot {\bf Q_{\bf
J}}$ vanish, however  $\bm \tau_{\rm STT}$ and $\bm\tau_{\rm lat}$
may both be non-zero, implying a coupling from the angular momentum
of the lattice to the magnetization.  This coupling flows from the
lattice to the orbital subsystem through the crystal field, which
then couples to the spin through spin-orbit coupling, and finally to
the magnetization through the exchange interaction.

\begin{figure}
\begin{center}
\vskip 0.2 cm
\includegraphics[width=2.75in]{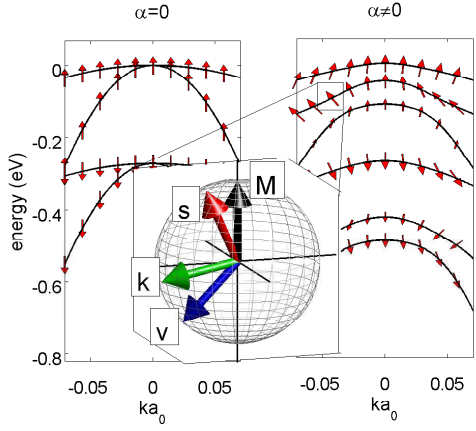}
\vskip 0.2 cm \caption{Left and right panels shows GaMnAs band
structure without and with spin-orbit, respectively (for
$\gamma_2=\gamma_3=2.4$). (arrows indicate spin direction of
eigenstates). The inset shows the direction of bulk magnetization,
and spin, velocity, and k vectors for a single state (in black, red,
blue, and green).  The torque from the misalignment between
magnetization and spin equals the torque from the misalignment
between velocity and k vectors.}\label{fig:bands}
\end{center}
\end{figure}

\noindent {\em Application to DMS} ---  We apply this general
formalism to a model of a dilute magnetic semiconductor (DMS).  DMSs
are semiconductor host materials which become ferromagnetic when
doped with magnetic atoms. ${\rm Ga_{1-x}Mn_x As}$ is the archetype
for these materials, and can be described as a system of local
moments of ${\rm Mn}$ $d$-electrons, whose interaction is mediated
by holes in the semiconductor valence band \cite{jungwirth}.  The
valence states are described by the Kohn-Luttinger Hamiltonian
$H_0^{\rm KL}$, which represents a small-${\bf k}$ expansion for a
periodic $H_0$, acting in the $\ell=1$ subspace (describing valence
states).  It is given by:
\begin{eqnarray}
H_0^{\rm KL} &=&
\frac{\hbar^2}{2m}\left(\left(\gamma_1+4\gamma_2\right) k^2 -
\frac{6\gamma_2}{\hbar^2} \left({\bf L} \cdot {\bf k}\right)^2
\right. \nonumber
\\&&\left.- \frac{6}{\hbar^2}\left(\gamma_3-\gamma_2\right)\sum_{i \neq j} k_i k_j L_i
L_j\right), \label{eq:KL}
\end{eqnarray}
where ${\bf L}$ are the spin-1 matrices for the $p$-state orbitals,
$\gamma_1,~\gamma_2,~\gamma_3$ are Luttinger parameters, and ${\bf
k}$ is the Bloch wave-vector.  Figure \ref{fig:bands} shows how the
presence of spin-orbit coupling affects the band structure.

For periodic systems the velocity operator can be written as: ${\bf
{\hat v}} = \frac{1}{\hbar} \frac{\partial H}{\partial {\bf k}}$,
and spin and angular momentum current densities are again defined as
symmetrized products of ${\hat {\bf v}}$ and ${\hat {\bf L}}$, and
${\hat {\bf v}}$ and ${\hat {\bf s}}$.  The dynamics of the
magnetization occur on a much longer time scale than that of the
electronic states, so we compute the dynamics from a sum over
scattering states, for which $\frac{d{\bf s}}{dt}=\frac{d{\bf
L}}{dt}=0$. For the Luttinger Hamiltonian, the $z$-component of
$\bm{\hat \tau}_{\rm lat}$ is:
\begin{eqnarray}
{\hat \tau}_{\rm lat}^z &=& \left({\hat {\bf v}}\times \hbar{\bf k}
\right)_z+\frac{6(\gamma_2-\gamma_3)}{\hbar
m}\left\{\left(k_xL_y+k_yL_x\right),\right.\nonumber\\
&&~~~~~~~~~~~~~~~~~~~~~\left.\left(k_xL_x-k_yL_y\right)\right\}~~~~
\label{eq:taulat}
\end{eqnarray}
where the brackets on the second term indicate an anticommutator.
Other components are given by cyclic permutation of indices. The
first term of Eq.~(\ref{eq:taulat}) can be written as ${\hat {\bf
v}} \times \hbar{\bf k} = \frac{d}{dt} \left({\hat {\bf r}} \times
\hbar {\bf k}\right)$. This term can be interpreted as a torque on
the crystal angular momentum ${\hat {\bf r}} \times \hbar{\bf k}$,
and results from the misalignment between wave vector and velocity.
It is generically nonzero for any material with a non-spherical
Fermi surface.  In the spherical approximation
($\gamma_2=\gamma_3$), Eq.~(\ref{eq:stt}) implies that the net flux
of total angular momentum into a volume is equal to the change of
magnetization plus the crystal angular momentum inside the volume.

\begin{figure}
\begin{center}
\vskip 0.2 cm
\includegraphics[width=2.75in]{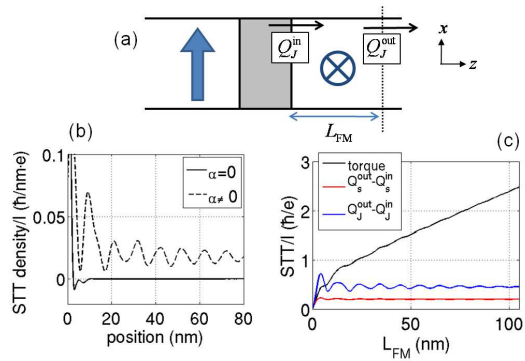}
\vskip 0.2 cm \caption{(a)Spin valve geometry: FM layers'
magnetization points in the $\hat x$ and $\hat y$ (out-of-page)
directions.  (b) The spin transfer torque versus position away from
the left normal metal-FM interface, which decays to zero in the
absence of spin-orbit coupling, and does not in its presence.  (c)
Plot of the total spin transfer torque, the net flux of spin
current, and net flux of total angular momentum current versus FM
thickness. The linear dependence for large thickness is due to a
persistent spin transfer torque.}\label{fig:tpl}
\end{center}
\end{figure}

\noindent {\em STT in spin-valves} --- We first consider a system to
study the $\bm\tau_{\rm STT}$ term of Eq.~(\ref{eq:stt}). Figure
\ref{fig:tpl}(a) shows the geometry; current flows in the $\hat
z$-direction, perpendicular to the magnetization of both layers. We
focus on the component of torque which is in the plane spanned by
the two magnetization directions.  This in-plane torque is
determined by the out-of-plane (or $\hat z$-component) spin density
\cite{nunez}. For the results presented here, we use the parameter
values: $(\gamma_1,\gamma_2,\gamma_3) = (6.85,2.1,2.9)$,
$\Delta=0.27~{\rm eV}$, $\alpha=0.11~{\rm eV}$, $E_{\rm F}=0.16~{\rm
eV}$ ($E_{\rm F}$ is measured from the top of the valence band). The
tunnel barrier is described by Eqs. (\ref{eq:HKL}) and
(\ref{eq:KL}), with $\Delta = 0$, and with an energy offset so that
the top of the valence band is 0.1 eV below $E_{\rm F}$.  We
calculate the eigenstates numerically and apply boundary conditions
as described in Ref. \cite{malik}.

Figure \ref{fig:tpl}(b) shows the spin transfer torque density as a
function of distance away from the interface.  We find that for
$\alpha=0$ (no spin-orbit coupling), the torque decays to zero away
from the interface, as expected \cite{stiles}.  For $\alpha \neq 0$,
the torque oscillates around a nonzero value, and extends into the
bulk. Figure \ref{fig:tpl}(c) shows that the total spin transfer
torque as a function of ferromagnetic (FM) layer thickness $L_{\rm
FM}$ is proportional to thickness for large $L_{\rm FM}$. This is in
contrast to the metallic spin valve, where the torque is an
interface effect and becomes constant for large $L_{\rm FM}$.

This persistent spin transfer torque arises because the spins of
individual eigenstates are not aligned with the magnetization (see
Fig.~\ref{fig:bands}) in the presence of spin-orbit coupling.  The
misalignment gives rise to a torque between the lattice and the
magnetization.  In equilibrium, these torques cancel when summed
over all occupied states.  However, the presence of a current
changes the occupation of the bulk states and can give rise to a
torque \cite{manchon,garate} in systems without inversion symmetry.
Inversion symmetry is only very weakly broken in bulk GaMnAs, and is
not included in the Kohn Luttinger Hamiltonian, Eq.~(\ref{eq:KL}).
Here, interfaces between materials breaks inversion symmetry.

The combination of an interface and a current flow changes the
occupation of the bulk states near the Fermi energy (depending on
the transmission probabilities of individual states across the
interface) and induces coherence between these states.  The change
in the occupation probabilities gives rise to a persistent
transverse spin accumulation, which only decays through other
scattering mechanisms not included here ({\it e.g.} defect
scattering). This spin accumulation gives rise to the persistent
spin transfer torque. The coherence between the states modifies the
spin accumulation and the torque near the interface but these
corrections decay away from the interface due to dephasing.

\begin{figure}
\begin{center}
\vskip 0.2 cm
\includegraphics[width=2.4in]{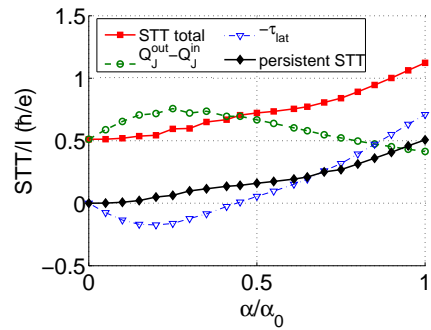}
\vskip 0.2 cm \caption{The total spin transfer torque, the net flux
of total angular momentum, $-\bm\tau_{\rm lat}$, and the persistent
component of the spin transfer torque on a FM layer with $L_{\rm
FM}=30~{\rm nm}$ as $\alpha$ is increased from 0 to
$\alpha_0=0.11~{\rm eV}$.}\label{fig:3}
\end{center}
\end{figure}

Figure \ref{fig:3} shows, as a function of the spin-orbit coupling
constant $\alpha$, the values of total spin transfer torque, the
angular momentum current flux, -$\bm \tau_{\rm lat}$, and the
persistent contribution to spin transfer torque (for $L_{\rm
FM}=30~{\rm nm}$). We determine the persistent contribution from the
slope of the integrated total versus FM width $L_{\rm FM}$ at large
$L_{\rm FM}$ (see Fig.~\ref{fig:tpl}c).  This procedure neglects the
contributions from coherence near the interface.  In this example,
the spin transfer torque increases with the addition of spin-orbit
coupling, largely because of the addition of the persistent term.
This qualitative behavior depends on system parameters: for $E_{\rm
F}=0.34~{\rm eV}$, for example, the spin-orbit coupling decreases
the total torque.

\begin{figure}
\begin{center}
\vskip 0.2 cm
\includegraphics[width=2.75in]{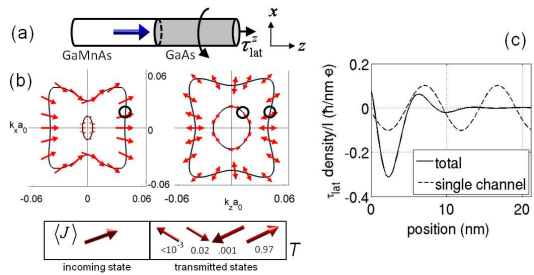}
\vskip 0.2 cm \caption{(a) shows the system geometry. We take
electron particle flow from left to right, and consider the
mechanical torque $\bm\tau_{\rm lat}$ in the z direction. (b) shows
slices of the Fermi surface for the different layers, with $\langle
{\bf J}({\bf k})\rangle$ superimposed, and also shows the ${\bf
J}$-character of the states specified by the black circle. Also
shown is the transmission probability for each of the states in the
GaAs. (c) shows the total mechanical torque density in the GaAs as a
function of distance from the interface (dark curve), and the
contribution from the single incoming state specified in (b) (dashed
curve). }\label{fig:4}
\end{center}
\end{figure}

\noindent {\em Nanomechanical torques in wires} --- We next consider
a system which exemplifies that physics of the $\bm\tau_{\rm lat}$
term of Eq.~(\ref{eq:stt}): a single interface between GaMnAs and
GaAs, with the direction of the magnetization parallel to the
current flow (see Fig.~\ref{fig:4}a).  This is similar to the
geometry considered in previous theoretical and experimental work
\cite{mohanty,kovalev,guiti}. The vanishing magnetization in GaAs
implies $\bm\tau_{\rm STT}=0$, so that $\bm\tau_{\rm
lat}=\nabla\cdot {\bf Q}_{\bf J}$, and its total value can be
deduced from ${\bf Q}_{\bf J}^{\rm in}-{\bf Q}_{\bf J}^{\rm out}$.
We use the same parameters as before, except $E_{\rm F}=0.06~{\rm
eV}$, and the top of the valence band of both layers coincide.

Figure \ref{fig:4}c shows $\bm\tau_{\rm lat}$ in the GaAs layer as a
function of distance away from the interface (assuming electron
particle flow from left to right).  The total torque (dark curve)
shows oscillatory decay, while the torque from a particular channel
(light curve) shows simple oscillation. The behavior of the single
channel is illustrated in Fig.~\ref{fig:4}b. We assume specular
scattering, so that the incident state chosen (black circle)
transmits into the four states of GaAs with equal $k_x,~k_y$ (also
shown with black circles). The character of these states, along with
the transmission probability, is shown in Fig.~\ref{fig:4}b.  The
incoming state couples most strongly to the state with similar ${\bf
J}$ character, but also partially transmits into other states with
different ${\bf J}$ character and wave vector $k_z$.  These
different scattering channels interfere with each other, leading to
an oscillatory ${\bf J}(z)$, with an oscillation period inversely
proportional to the splitting of $k_z$ wave-vectors of the different
sheets of the Fermi surface.  This splitting is from the lattice
crystal field and spin-orbit coupling, the agents responsible for
$\bm\tau_{\rm lat}$.  Different channels have different oscillation
periods, so that their total decays away from the interface, as
happens for spin transfer torques in ferromagnets \cite{stiles}. For
the parameters used here, we find $Q_{J^z}^{\rm in} = 1.20 \hbar
\frac{I}{e}$, due to the polarization of the states from the
magnetization, while $Q_{J^z}^{\rm out} = 0.46 \hbar \frac{I}{e}$.
Mechanisms not considered here, such as spin-flip scattering, ensure
that $Q_{J^z}^{\rm out}$ decays to zero away from the interface.

The mechanical torque is $Q_{J^z}^{\rm in} - Q_{J^z}^{\rm out} =
0.74 \hbar \frac{I}{e}$.  For appropriate experimental conditions,
this torque is greater than the thermal fluctuations and is a
measurable effect. We refer the reader to Ref.
\cite{mohanty,kovalev,guiti} for details of treatment of the torsion
dynamics and experimental details.  The formalism developed here
generalizes previous work to allow for microscopic evaluation of the
electronic structure contribution to the current-induced mechanical
torque. For systems with nonzero magnetization, the microscopic form
of $\bm \tau_{\rm lat}$ is necessary to determine the partitioning
of total angular momentum flux between torques on the magnetization
and torques on the lattice. Our theory neglects other mechanisms of
spin relaxation, such as disorder-induced spin-flip scattering, so
that full calculations will require microscopic calculations like
these to be embedded in diffusive transport calculations.

\noindent {\em Conclusion}--- We have shown how atomic-like
spin-orbit coupling affects current-induced torques: both the spin
transfer torque on the magnetization and the mechanical torque on
the lattice.  In GaMnAs spin valves, we find a contribution to the
spin transfer torque that persists throughout the bulk. This result
may explain experiments which find critical currents which are up to
an order of magnitude smaller than the value expected from a simple
accounting of the net spin current flux \cite{chiba,elsen}. For a
single interface between GaMnAs and GaAs, we microscopically compute
the mechanical torque due to scattering from the interface. These
results highlight important, qualitatively different physics at play
when spin-orbit coupling is strong.

The authors acknowledge helpful conversations with A. H. MacDonald.


\begin{thebibliography}{00}
\bibitem{bergerdw}
L. Berger, J. Appl. Phys. {\bf 3}, 2156 (1978); ibid. {\bf 3}, 2137
(1979).

\bibitem{slonc}
J. Slonczewski, J. Magn. Magn. Mat. {\bf 62}, 123, (1996).

\bibitem{berger}
L. Berger, Phys. Rev. B {\bf 54}, 9353 (1996).

\bibitem{ralph}
D. C. Ralph and M. D. Stiles, J. Magn. Magn. Mater. {\bf 320}, 1190
(2007).

\bibitem{miltat}
M. D. Stiles and J. Miltat, Top. Appl. Phys. {\bf 101}, 225 (2006).

\bibitem{katine}
J. A. Katine and E. E. Fullerton, J. Magn. Magn. Mater. {\bf 320},
1217 (2007).

\bibitem{richardson}
O. W. Richardson, Phys. Rev. {\bf 26}, 248 (1908).

\bibitem{einstein}
A. Einstein and A. de Hass, Verhandlungen der Deutschen
Physikalischen Gesellschaft, {\bf 17}, 152 (1915).


\bibitem{mohanty}
P. Mohanty {\it et al.}, Phys. Rev. B {\bf 70}, 195301 (2004).

\bibitem{kovalev}
A. A. Kovalev {\it et al.}, Phys. Rev. B {\bf 75}, 014430 (2007).

\bibitem{guiti}
G. Zolfagharkhani {\it et al.}, Nature Nanotech. {\bf 3}, 720
(2008).

\bibitem{footnote1}
In addition to atomic-like angular momentum, there is a contribution
to the total orbital angular momentum from itinerant motion through
the lattice. The distinction between ``local" and ``itinerant"
orbital angular momentum is discussed in Ref. \cite{vanderbilt} . In
this work, we consider only the atomic-like contribution.

\bibitem{vanderbilt}
T. Thonhauser {\it et al.}, Phys Rev. Lett. {\bf 95}, 137205 (2005).

\bibitem{nunez}
A. S. N\'{u}\~{n}ez and A. H. MacDonald, Solid State. Comm. {\bf
139}, 31 (2006).

\bibitem{stiles}
M. D. Stiles and A. Zangwill, Phys. Rev. B {\bf 66}, 014407 (2002).

\bibitem{footnote2}
For $H_0 = -\hbar^2\nabla^2/2m+V({\bf r})$, our definition of
$\hat\tau_{\rm lat}$ is equivalent to $\left[V({\bf r}),\hat {\bf
L}\right]$.  We use $\hat\tau_{\rm lat} = \left[H_0,\hat {\bf
L}\right]$ in anticipation of other forms of $H_0$, in particular
the $\bf k\cdot \bf p$ form of the Luttinger Hamiltonian.

\bibitem{jungwirth}
T. Jungwirth {\it et al.}, Rev. Mod. Phys. {\bf 78}, 809 (2006).

\bibitem{malik}
A. M. Malik {\it et al.}, Phys. Rev. B {\bf 59}, 2861 (1999).

\bibitem{manchon}
A. Manchon and S. Zhang, Phys. Rev. B {\bf 78}, 212405 (2008).

\bibitem{garate}
Ion Garate and A. H. MacDonald, Phys. Rev. B {\bf 80}, 134403
(2009).

\bibitem{chiba}
D. Chiba {\it et al.}, Phys. Rev. Lett. {\bf 93}, 216602 (2004).

\bibitem{elsen}
M. Elsen {\it et al.}, Phys. Rev B {\bf 73}, 035303 (2006).


\end{thebibliography}
\end{document}